# Angular Dependence of Four Mechanical Properties of Graphynes

Guilherme B. Kanegae and Alexandre F. Fonseca*

*Applied Physics Department, Gleb Wataghin Institute of Physics, University of Campinas - UNICAMP, Campinas, São Paulo, CEP 13083-859, Brazil.*

*Corresponding author: afonseca@ifi.unicamp.br

## ABSTRACT

*Graphyne is a porous two-dimensional carbon allotrope of graphene that possesses interesting physical properties, including non-null bandgap. It is composed of carbon hexagonal rings or carbon-carbon bonds connected by acetylenic chains. The diverse forms of these connections yield a variety of graphyne structures. In a previous study, we have obtained the elastic properties of seven distinct families of graphyne structures as a function of the number of acetylene chains, from 1 to 10. The Young's modulus, shear modulus, Poisson's ratio and linear compressibility were predicted for the zigzag and armchair directions of all 70 graphyne structures. Here, we present some noteworthy findings regarding the angular dependence of these four elastic properties of asymmetric graphynes. Our results demonstrate that in a single structure, the minimum and maximum Young's modulus can vary by a factor of 10. Additionally, the directions of null linear compressibility in some asymmetric structures were determined.*

## INTRODUCTION

Graphyne (GY) is a two-dimensional porous carbon nanostructure. The structure was first proposed by Baugham, Eckhardt, and Kertesz in 1987 [1]. It is formed by connecting hexagonal rings or carbon-carbon bonds through acetylene chains of length $n$, $(-[-C\equiv C-]n-)$ [2]. It has been shown in the literature that GYs exhibit good electronic, mechanical, and thermal properties [3-8], including a non-zero band gap [1,9], which is a desirable attribute for applications in electronics. The first synthesis of a GY with $n = 2$ was reported in 2010 [10]. Subsequently, other researchers were able to synthesize it [11,12], as well as synthesize GYs with $n = 4$ [13,14]. The synthesis of GYs with $n = 1$ has only recently been reported [15-17], generating excitement about this promising two-dimensional carbon nanostructure. The porous nature of GYs has been demonstrated to be a special attribute with potential applications ranging from hydrogen storage to antibacterial agents [18-21].

In a recent study [22], we conducted a combined theoretical and computational investigation of the elastic properties of 70 GY structures, i.e., with number $n$ of acetylenic chains varying from 1 to 10 for each of the original seven families of GYs [1]. Computational techniques were employed to calculate the elastic coefficients ($C_{ij}$) and, subsequently, the Young's modulus, shear modulus, Poisson's ratio and linear compressibility of all 70 structures. In addition, a mechanical model based on a serial association of $n$ springs was shown to accurately represent the dependence on $n$ of the elastic properties of GYs, with the exception of the Poisson's ratio. Furthermore, we expanded the analysis to study the density dependence of the elastic properties of GYs, demonstrating that the elastic modulus of GYs exhibits a lesser dependence on density, $\rho$, than that observed in porous cellular materials [23]. Here, we present the angular dependence of all four elastic properties of GYs. Our results show that asymmetric GYs exhibit remarkable attributes along several different directions. We will discuss our findings in terms of the general contributions to the elasticity of a structure from stretching and hinging.

For the sake of simplicity, each GY is named as "G$n$Y$f$", where $n$ ($f$) is the number of acetylene chains, with $1 \leq n \leq 10$ (the corresponding GY family or type, with $1 \leq f \leq 7$), according to Ivanovskii notation [2]. Figure **1** shows the structures of the 7 GY families with $n = 1$.

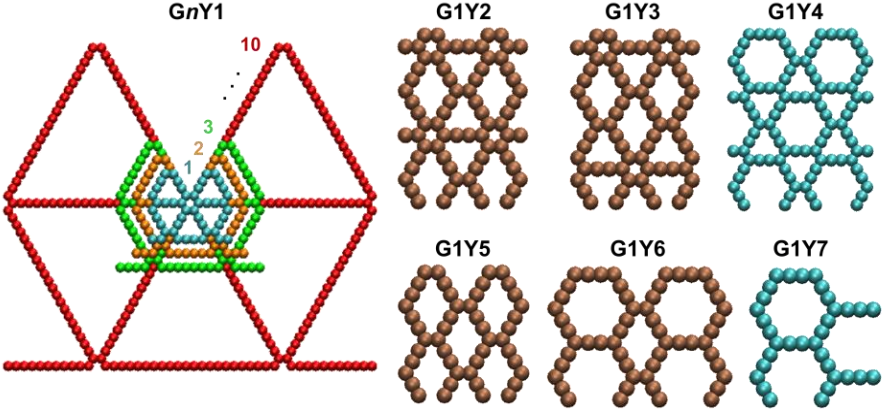

**Figure 1. Left**: superposition of the graphyne structures belonging to the first family, G$n$Y1, with different numbers of acetylenic chains, $n$. The cyan, orange, green and red colors correspond to $n = 1, 2, 3$ and $10$, respectively. **Right**: the $n = 1$ members of the other six families of GYs as originally proposed by Baughman, Eckhardt and Kertesz [1]. For all structures, the GY's armchair and zigzag directions are drawn along the horizontal and vertical directions, respectively. The numbering notation is the same as defined in Ref. [22]. Symmetric (asymmetric) structures are drawn in cyan (brown).

In the next sections, the theory and computational methods are described, as well as, the results and the discussion. At the end, we summarize the conclusions.

**THEORY AND SIMULATION DETAILS**

The computational simulations employed to obtain the in-plane elastic constants $C_{ij}$ of GYs followed the protocols given in Ref. [22]. The LAMMPS [24] package and AIREBO potential [25] were used to obtain the equilibrium structures. The Young's modulus, $E$, the shear modulus, $G$, the linear compressibility, $\beta$, and the Poisson's ratio, $\nu$, of all GYs, along the armchair direction, were calculated using the following equations:

$$E_x = \frac{C_{11}C_{22} - C_{12}^2}{C_{22}},\ G = C_{66} = \frac{1}{4}(C_{11} - 2C_{12} + C_{22}),\ \nu_{xy} = \frac{C_{12}}{C_{22}} \text{ and } \beta_x = \frac{C_{22} - C_{12}}{C_{11}C_{22} - C_{12}^2}. \quad (1)$$

These methods and equations have been employed in analogous studies of the mechanical properties of graphynes and other two-dimensional structures [26,27].

In order to elucidate the angular dependence of $E$, $G$, $\beta$ and $\nu$ for all GY structures, we rotate the elastic constant matrix, $\mathbf{C}'$, with the following equation:

$$\mathbf{C'} = \mathbf{R}(\theta) \cdot \mathbf{C} \cdot \mathbf{R}(\theta), \qquad (2)$$

where $\mathbf{R}(\theta)$ represents a rotation matrix with an angle of $\theta$. The equations (1) are employed to determine the values of $E$, $G$, $\beta$ and $\nu$ along a direction that forms an angle of $\theta$ with the original $x$-axis or horizontal (armchair) direction.

It is anticipated that the elastic properties of symmetric GYs (families G$n$Y1, G$n$Y4, and G$n$Y7) remain constant with respect to $\theta$. Nevertheless, it can be demonstrated through a relation between $C_{66}$ and the remaining $C_{ij}$ [28], that the shear modulus of all structures, symmetric or not, is independent of $\theta$. To illustrate that, Figure **2** shows the calculated Young's modulus of all ten members of the symmetric G$n$Y1 family and the shear modulus of all members of the asymmetric G$n$Y5 family.

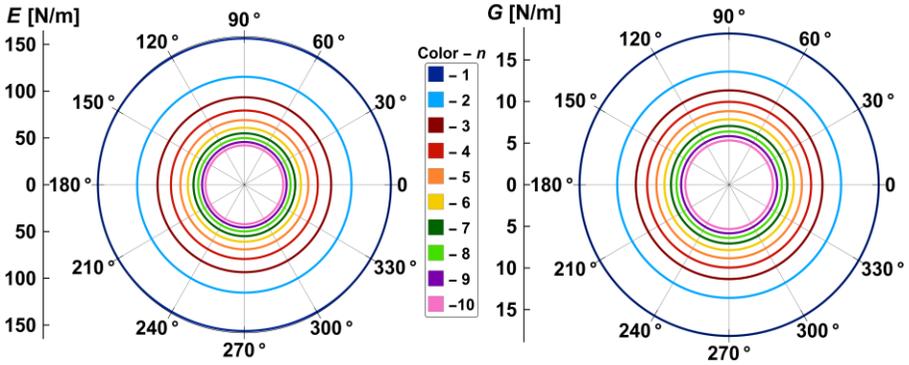

**Figure 2.** Polar plots of the Young's modulus (left panel) of the symmetric G$n$Y1 family and the shear modulus (right panel) of the asymmetric G$n$Y5 family. The line colors correspond to structures with varying numbers of acetylene chains.

## RESULTS AND DISCUSSION

As anticipated, the most intriguing findings emerge from the asymmetric GYs, namely families G$n$Y2, G$n$Y3, G$n$Y5 and G$n$Y6. Figure **3** shows the Young's modulus, Poisson's ratio and linear compressibility of all GY structures from families G$n$Y3, G$n$Y5 and G$n$Y6. The outcomes for family G$n$Y2 are not so different from those of G$n$Y3 and are thus not depicted to conserve space.

The Young's modulus of the GY structures from families G$n$Y5 and G$n$Y6 is found to be significantly more angular dependent than that of families G$n$Y2 and G$n$Y3. These structures exhibit larger modulus along the directions of the acetylenic chains (60°) than along zigzag and armchair ones. This is due to the fact that carbyne, which is an infinite acetylene chain, is one of the structures with the highest tensile modulus [29]. The G$n$Y5 structures exhibit the largest maximum to minimum Young's modulus ratio (approximately 10, from values as low as approximately 4 N/m along armchair direction to approximately 40 N/m along $\theta = 60°$). A tensile stress applied along 0° or 90° direction induces hinging movements at the acetylenic chain junctions. At 60°, however, the tensile stress will be felt by the acetylene chain, itself. A similar analysis can be conducted for the other families, with the additional observation that the G$n$Y3 and G$n$Y6 structures possess acetylene chains in armchair directions, which decreases their maximum to minimum Young's modulus ratio. A noteworthy qualitative result is that as $n$ increases, the symmetry of the polar plot curves decreases. This phenomenon can be explained by considering two contributions to the elasticity of the structure: stretching and hinging strains. According to

Grima [30], stretching and hinging play different roles in the determination of the elastic properties of a structure. As the number of acetylene chains increases, the hinging factor predominates, increasing the asymmetry of the polar plot curves.

The polar plots of the Poisson's ratio of the asymmetric GY structures are also asymmetric. As for the Young's modulus, they are large (small) along zigzag (armchair) directions. The greatest asymmetry in the Poisson's ratio is observed in the G$n$Y5 structures. In contrast with the Young's modulus, where larger values of $n$ correspond to smaller values, a larger the value of $n$, larger the Poisson's ratio. This result corroborates the hypothesis that as $n$ increases, the hinging factor becomes more influential in the elastic response of the GY structures. As shown by Grima [30], the pure stretching of honeycomb and wine-rack structures results in negative Poisson's ratio. Consequently, as $n$ increases, the contribution of stretching to the Poisson's ratio decreases.

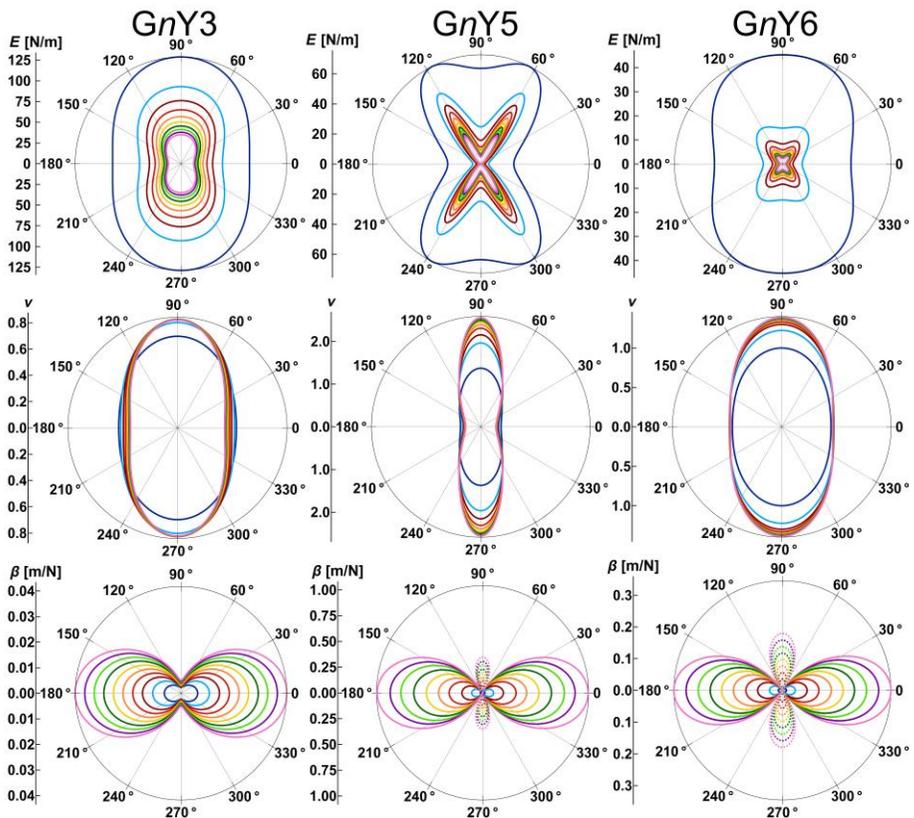

**Figure 3.** Young's modulus (top), Poisson's ratio (middle) and linear compressibility (bottom) of GY structures from G$n$Y3, G$n$Y5 and G$n$Y6 families. Positive and negative values of the linear compressibility of GYs from G$n$Y5 and G$n$Y6 families are drawn in full and dashed lines, respectively. The color code is the same as in Figure **2**.

The linear compressibility of the asymmetric GYs displays a broad range of outcomes. As with Poisson's ratio, the maximum linear compressibility increases with $n$. The ratios of the maximum to minimum linear compressibility for the G$n$Y2 and G$n$Y3 families are approximately 4 and 10, respectively. However, the most interesting findings are observed in the G$n$Y5 and G$n$Y6 families. In contrast to the G$n$Y2 and G$n$Y3 families, for which the linear compressibility is consistently positive, the G$n$Y5 and G$n$Y6 families exhibit a reversal in sign, resulting in negative linear compressibility. The negative values of linear compressibility are indicated by dashed lines in Figure **3**. The maximum absolute values of the positive linear compressibility of the G$n$Y5 and G$n$Y6 families are greater than the maximum absolute values of the negative ones. The coexistence of positive and negative linear compressibility in different directions implies the existence of directions along which the linear compressibility of the GY structures is null. The polar plots indicate that these directions are along values that are multiples of 60°. An alternative representation of these results is provided by a contour plot. Figure **4** depicts two contour plots of the linear compressibility, β, of GYs of G$n$Y5 and G$n$Y6 families. The red (blue) regions indicate positive (negative) values of the linear compressibility while white regions represent zero values. As evidenced by the color intensity, the absolute values of linear compressibility increase with $n$. Additionally, the contour plot highlights the null linear compressibility directions at 60°, 120°, 240° and 300° angles.

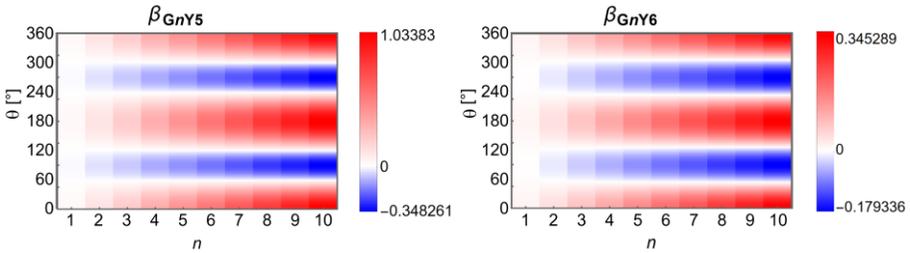

**Figure 4.** Contour plots of the dependence of the linear compressibility, β, of G$n$Y5 (left) and G$n$Y6 (right) structures on both θ and $n$. Red and blue regions correspond to positive and negative values of linear compressibility, respectively.

## CONCLUSION

In this study, we presented the angular dependence of four elastic properties of 70 GY structures. With the exception of the shear modulus, the Young's modulus, Poisson's ratio and linear compressibility were shown to be angular dependent in asymmetric G$n$Y2, G$n$Y3, G$n$Y5 and G$n$Y6 structures. It was observed that the ratio of maximum to minimum Young's modulus and linear compressibility can reach as much as 10 for some GY structures. The G$n$Y5 family exhibited the most interesting properties, from the ability to present an as high Poisson's ratio as approximately 2.5 to exhibiting null or negative linear compressibility. We have discussed and interpreted these results in terms of the balance between stretching and hinging of the acetylene chains.

## ACKNOWLEDGMENTS

This work used resources of the John David Rogers Computing Center (CCJDR) in the Gleb Wataghin Institute of Physics, University of Campinas.

## FUNDING

This work was supported by the Coordenação de Aperfeiçoamento de Pessoal de Nível Superior - Brasil (CAPES) - Finance Code 001; the Brazilian Agency CNPq-Brazil (Grant number 303284/2021-8); São Paulo Research Foundation (FAPESP) (Grant numbers #2020/02044-9 and #2023/02651-0); and Fundo de Apoio ao Ensino, Pesquisa e Extensão – FAEPEX/UNICAMP (Grant number #3072/24).

## AUTHOR CONTRIBUTION

A.F.F. conceived the idea and G.B.K. performed the computational simulations and calculations. A.F.F. and G.B.K. analyzed the results. A.F.F wrote the original draft and G.B.K. reviewed and edited the draft. A.F.F supervised the work and acquired funding. A.F.F. and G.B.K. read and approved the final manuscript.

## CONFLICT OF INTEREST STATEMENT

On behalf of all authors, the corresponding author states that there is no conflict of interest.

## DATA AVAILABILITY STATEMENT

Data available on reasonable request from the authors.